# Two-Group Flux Analysis of Neutrons which enter, internally scatter, and often escape a Shielding Layer of Iron- with respective Group Settings circa one MeV and kilo-eV

Eric V Steinfelds[a,b], Keith Andrew[a]

[a] Department of Physics and Astronomy, Western Kentucky University, Bowling Green, KY
[b] Alverno College, Milwaukee, WI

Email: *eric.steinfelds@wku.edu*, *keith.andrew@wku.edu*, *eric.steinfelds@alverno.edu*,

**Abstract**

It has been long recognized that radiation transport theory is the foundation for the planning and analysis of X-ray (γ-ray) radiation therapy and for imaging. In less common but appropriate occasions as an alternative to X-rays or gammas, neutron radiation is used in oncological treatments or in imaging of patients. The following work is also potentially of interest to Radiation Safety Planners. Especially in regard to uses for neutron beams, we introduce and present a deterministic and semi-analytical method for doing transport analysis on neutrons and which are judiciously set to be distributed into two energy groups. There are advantages for doing such 2-group and higher multi-group analysis of radiative particles (i.e. neutrons and photons). These advantages are that we can more directly keep track of what percentages of radiative particles are close to the original high energy and how many are at significantly lower energy. For photons, the profile of any build-up function shows that the function is slightly larger than 1.0 at entry, then it rises to perhaps 2 or 3 within roughly one mean free path of the fast primary particles, and finally approaches the asymptote of 1.0 as the penetration depth gets progressively larger. Neutrons deserve a separate treatment. Although it is lengthier, our algorithm and formulation is much more complete than the popular formula used among radiologists of Intensity($x$) = $B(E, x)$ Intensity(0) $exp(-\mu \cdot x)$. Moreover, a buildup function for neutron fluxes do not appear to be widely offered in radiological and radiation safety publications. This paper show predicts the ratio of high energy (circa 1 MeV) neutrons over low energy neutrons (in a group below 0.201 MeV) which are scattered backwards and forwards out of walls of Fe-56 at various thicknesses.

**Introduction**

The field of health physics, the essential background topics in medical physics, and the reviews of radiological safety of reactors of nuclear reactor operations have been served by various methodically practiced styles of dosimetry/transport calculations for over 70 years. This has been done with clinical caution and practical verifiability in mind in the main accepted traditions in the disciplines of health physics and in "reactor physics" engineering. Radiation shielding calculations, penetration assessments[1], and radiation dosimetry calculations[2] are included among the vital and materially informative calculations. Tying into the physics of interaction of radiation with matter, a beam which is comprised of either energetic neutrons or high-energy photons (i.e., gamma rays or hard X-rays) have conventionally been looked upon as candidates for interception via the standard X-section (i.e., cross-section) inspired models, for which one uses the typical expression

$$I(z) = I(0)e^{-\mu z} \qquad (1)$$

where z is the depth and μ is the attenuation parameter or 'coefficient'.

The "simplistic" attenuation formula is an appropriate name for this short Eq. (1). Many shielding calculations done by medical physicists and health physicists over the years (since at least 1970) have been done using an extension of this "simplistic attenuation formula" coupled with a buildup factor[3]. The buildup factor is necessary if we are to use an attenuation formula as our principal tool of analysis to account for scattered neutrons or energetic photons, respectively[4]. Some of these "n's" and "γ's" are almost (or totally) elastically scattered, but some of these are down-scattered in energy. In keeping with the aspiration for excellence from the era of the "Space Age", some computer/electrical engineers and nuclear engineers have enhanced these efforts by conducting Monte Carlo simulations of the transport and penetration of neutrons or photons through walls and various barriers with various respected codes (packages with X-section libraries) such as MCNP, EGS4, EGSNrc, and the versatile but highly tedious GEANT4[5,6,7]. In this short paper, we introduce and present a deterministic and semi-analytical method for doing transport analysis on neutrons and isotropically scattering 'hard' photons which are placed in two energy and, with future ambitions, into 3 energy groups. There are advantages for doing such 2-group and higher multi-group analysis of radiative particles (i.e., neutrons and photons). These advantages are that we can more directly keep track of what percentage of radiative particles are close to the original high energy and how many are at significantly lower energy. An inspection of the profile of any build up function shows that the function is slightly larger than 1.0 at entry, then it rises to perhaps 2 or 3 within roughly one mean free path of the fast primary particles, and finally approaches the asymptote of 1.0 as the penetration depth gets progressively larger.

**Review of buildup and discussion of methods of analysis**.

The attenuation formula in the introduction expresses the particulate intensity, not the energetic intensity in our convention. This corresponds to the choice of analysis of particulate flux rather than energetic flux of radiation. For those with a non-nuclear background: "Flux" is used by the medical physics community and health physicists with a definition considerably different from that of the flux of electric fields. Our flux has units of "particles" per $cm^2$ per second. See Frank Attix's text[8] if this is unclear. Particulate intensity is less than or equal to the scalar flux of the particles. Very often intensity is defined as the magnitude of net current of transported particles per $cm^2$ of surface per sec. Indeed, in a case where equally many particles approach and penetrate a wall bidirectionally the net current is zero. However, the scalar flux in such an example is much larger than zero. Admittedly, one can make some inferences on the approximate ratio of down-scattered particles at a given depth as a function of position by inspecting the build up function, should it be available in published tables for a given shape and material. Here the intensity with a buildup coefficient can be expressed as

$$I(z) = B(E,z) I(0) e^{(-\mu z)} \qquad . \qquad (2)$$

However, the Buildup coefficient (i.e. $B(E, z)$) does not make clear just what percentage of scattered γ-ray or X-ray radiation is scattered so as to retain most of its energy and how much has been "demoted" to photons with 50% less or more of energy per radiative particle. This reality (of radiative particles often undergoing elastic or nearly elastic scattering) holds for 'free' neutrons which scatter off nuclei with an Atomic Number greater than 4. For example, when a 'n' with a kinetic energy of 1.0MeV collides with an Fe-56 nucleus, it has a 99% chance of undergoing elastic or quasi-elastic scattering (retaining most of its 1MeV). This same 'n' has a probability of less than 0.8% of down-scattering to a "low" energy neutron with less than 0.201MeV of kinetic energy. The chance of capture at 1MeV is less than one in a 1000. Fast neutrons are not easily captured, they usually are just scattered. This information is given based on inspecting generally available public claims of two group data and, more importantly, by our careful conducting of MCNP[9] simulations in which we reproduce the conditions of scattering and so called "buildup" of scattered neutrons in rectangular walls of Iron (Fe-56). We are authorized to use and have extensive experience with the very versatile MCNP Monte Carlo code. Regarding neutrons, the buildup coefficient data either is not widely published for nuclear engineers or not readily available. Thus, the case of neutron shielding analysis offers a major service for performing multi-group energetic neutron flux and dose calculations. In this paper we stick with 2-group n's (i.e., neutrons). The following formula is an overly simplistic and inadequate expression often used for metals with atomic number less than 84, away from the "uranic" family

$$\Phi(z) = \Phi(0) e^{(-\Sigma_{[totl, at, 1MeV]} \cdot z)} \qquad . \qquad (3)$$

However, Eq. (3) gives a very incomplete story of the local neutron flux. In analogy to fluxes of photons, Eq. (1), in which $I(z)$ is set to equal $I(0) \cdot \exp(-\mu z)$, also gives an incomplete story of local current/intensity of X-rays and γ-rays. This occurs in the equations above when one fails to include the Buildup factor. The Buildup factor is included in Eq. (2) as the expression B(E, z). B(E,z) corrects the local 'flux' of γ-rays or photons at various depths (e.g. z) of penetration. It is more insightful to replace I(z) from Eq. (1) with Φ(z) in Eq. (3) where flux is more appropriate than 'n' or photon current in the geometry of a wall or box since many of the neutrons or photons no longer travel straight forward along the z-axis after one or two collisions of scatter occur.

If the nuclei/atoms of a medium which is entered by the beam of n's or photons is a pure absorber, then Eq. (1) and Eq. (3) are acceptable solutions for penetration and the differential equation which explains the transport of the particle(s) is given by:

$$\frac{d(I_{[2]}(z))}{dz} + (\mu)I_{[2]}(z) = 0, \quad \text{for } \mu = \sigma_{absorb} n_{rad} \quad (4)$$

and where $n_{rad}$ is the number density from radiation. If the medium is Boron-10 and the neutrons are at low energy, then Eq. (4) would be a realistic equation for modeling transport of the neutrons, because B-10 is nearly a pure absorber. However, most materials are not pure absorbers.

If we presume that there are two energy levels for neutrons and photons (i.e., fast n's and slow n's), then it is appropriate to write a double energy-group Maxwell Boltzmann Transport Equation (MBTE) in order to express what is going on for transmission and for energy demotions (i.e., down scatters). This pair of equations is written below for the transport of neutrons with two possible energy levels[10,11,12]. Note in Eqs. (5) that $\Psi_{[2]}$ is proportional to $1/(4\pi)\cdot\Phi_{[2]}$ for fast neutrons (of grp-2). Here $\Phi_{[2]}$ is the scalar flux which includes neutrons in the energy grouping of 0.201MeV up to 10MeV (at least when we model and analyze iron shielding). $\Phi_{[1]}$ is the scalar flux which includes all neutrons of energy 0.20MeV and lower. Here is the two-group MBTE:

$$\vec{n}_{(\theta)} \cdot \vec{\nabla}\Psi_{[2]}(\vec{r},\theta) + (\Sigma_{[ab,2]} + \Sigma_{[s,2]} + \Sigma_{s[2,1]})\Psi_{[2]}(\vec{r},\theta) = \oint \Sigma_{[s,2]} \cdot \Psi_{[2]}(\vec{r},\theta') \cdot (_{jacobi(\angle)}) d\theta' \quad (5a)$$

$$\vec{n}_{(\theta)} \cdot \vec{\nabla}\Psi_{[1]}(\vec{r},\theta) + (\Sigma_{[ab,1]} + \Sigma_{[s,1]})\Psi_{[1]}(\vec{r},\theta) = \oint \left(\Sigma_{[s,1]} \cdot \Psi_{[1]}(\vec{r},\theta') + \Sigma_{s[2,1]} \cdot \Psi_{[2]}(\vec{r},\theta')\right) \cdot (_{jacobi(\angle)}) d\theta' \quad . \quad (5b)$$

Suppose that $\Sigma_{[s,2]}$ and $\Sigma_{s[2,1]}$ are equal to zero, as we might imagine for a medium of "super-Boron". Then, Eq.(5a) can be rewritten as: $\vec{n} \cdot \vec{\nabla}\Psi_{[2]}(\vec{r},\theta) + (\Sigma_{[ab,2]}) \cdot \Psi_{[2]}(\vec{r},\theta) = 0$. If our wall is very broad and if the distribution to approximation depends only on coordinate z, then this simple 1st order equation is completely equivalent to Eq. (4) with attenuation. In this paper, we presume that scattering is isotropic, which is often a good approximation for the scattering of neutrons. Therefore, we establish that $\Sigma_{[s,2]}$ and $\Sigma_{s[2,2]}$ are completely independent of $\theta$, $\theta$`, and any angle. This 2-group version of the MBTE is an example of a pair of integro-differential equations. Generally, it is easier to solve a purely integral equation, such as a Fredholm Int. Equation[13]. Holding on to the presumption of isotropic scattering, Eqs. (5a) and (5b) can be subjected to a special integral transformation via Green's functions in order to re-express them as the following integral equations,

$$\Phi_{[2]}(\vec{r}) = \frac{1}{4\pi} \oiiint \frac{1}{|\vec{r}-\vec{r}`|^{\wedge 2}} \left(\Sigma_{s_{2,2}} \cdot (\Phi_{[2]}(\vec{r}`) + I_{beam}(\vec{r}`))\right) e^{(-\Sigma_{[totl,2]} \cdot (\vec{r}-\vec{r}`))} d\vec{r}` \quad (6a)$$

$$\Phi_{[1]}(\vec{r}) = \frac{1}{4\pi} \oiiint \frac{1}{|\vec{r}-\vec{r}`|^{\wedge 2}} \left(\Sigma_{s_{1,1}} \Phi_{[1]}(\vec{r}`) + \Sigma_{s[2,1]} \Phi_{[2]}(\vec{r}`)\right) e^{(-\Sigma_{[totl,1]} \cdot (\vec{r}-\vec{r}`))} d\vec{r}` \quad . \quad (6b)$$

Eqs. (6a) & (6b) are challenging to solve, but some solutions have been found both by the Russian Math academy of the 1950s and by a 'computations' group at Los Alamos in the 1950s. The author(s) have found a method for numerically, with arbitrary precision, to iteratively solve the monoenergetic version of Eq.(6a) and the 2-group Eqs. (6)[14].

Eqs. (6) are more demanding and difficult to solve than it is to simplistically use Eq. (2) with the buildup coefficient to calculate relative intensities. The information which we get out of Eqs. (6) for $\Phi_{[2]}$ and $\Phi_{[1]}$ is much richer than what we can get from the calculation of I(z) as a function of penetration into a wall Eq. (2). This is especially apparent for a very broad slab of shielding (presuming breadth of slab is more than 6 times the thickness of slab). Presuming a beam of neutrons enters from the left at the interface where z=0, the user would need to read, download, and interpolate a table of build-up coefficient values (or use an approximate local formula) for a slab of the given material (such as Fe, Pb, or concrete). These tables are almost non-existent for neutrons. Such tables do exist for γ-ray and X-ray photons for industrial and medical materials, but they are limited in their range and diversity of examples. Thus, the dosimetrist who is borrowing or using the data often is stuck with having to interpolate from near fits of other examples most similar to the geometry which he or she has chosen to design or assess for predictions or dose verifications. As a reminder of geometrical concepts, the portion of "battered" particulate current density which escapes from the right-hand boundary of a rectangular slab of shielding from "mid-face" equals approximately ½ or 0.6 times the "battered" scalar flux which is present on the boundary of escape from the rectangularly shaped shield. 'Battered' shall be defined as the condition of a particle which has either been coerced or forced to scatter so as to keep some or all of its kinetic energy. "Battering" of a neutron changes its direction and can reduce its energy. For a very thin shield whose thickness is less than ⅛ of a mean free path of a neutron, the first author has verified that $I_{['battered']}$(escape)= ½ $\Phi_{['battered']}$(boundary) by analytical means of integrating the Green's function of the radiative source and by consideration of Gauss's Law with a discrepancy of less than 2 percent in lieu of rare double-scatter histories. For thicker samples MCNP simulations have been verified with converging persuasiveness that the escape ratio is in the range of ½ through 0.65, depending on thickness of shielding. $I_{['battered']}$ refers to the particulate current density of n's which have undergone at least one collision before escaping from either the left-hand side (LHS) or right-hand side (RHS) of the slab. Many books casually refer to the 'I' in $I_{['battered']}$ and $I_{[entering, beam]}$ as intensity. In this paper, 'I' shall mean current density of particles (mostly neutrons here) per cm² (or m²) per second. If the beam and scattered particle are all mono-energetic, then, of course, energetic I, or energetic intensity, is given by the product of 'I'(of 'n') and Energy$_{[of 'n']}$.

**Computations and Results**

Following our earlier work[15], we can iteratively solve integral equations (6a) and (6b) in the case where we have one very broad rectangular slab. If there is a monoenergetic beam of fast neutrons which enter the designated slab of shielding, then before the first iteration we find that $\Phi_{[2][0]}(z)$ goes as $I(0)e^{(-\Sigma_{[total,1MeV]}z)}$, as in Eq. (3) and where I(0) and $\Phi[0]$ were interchanged. $\Phi_{[2][0]}(z)$ is that part of the flux of neutrons at z that have never undergone scattering. The first index, which holds [2], shall denote that this is flux/current of the fast group of neutrons. In the first iteration (upon the event of scatter) we find the formula of $\Phi_{[n][1]}(z)$ for the n-group's neutrons where n equals either 1(slow) or 2(fast). Upon integration this flux can be found analytically for the n-group of neutrons. Consider Eq. (7)

$$\tilde{\Phi}_{[2]}(\vec{r}) = I(0)e^{-\left(\Sigma_{[total,1MeV]}z/\cos(\theta)\right)} + \Phi_{[2][1]}(z) + \Phi_{[2][2]}(z) + \Phi_{[2][3]}(z) + ... \qquad (7)$$

where $\Phi_{[2][n]}(z)$ is generated iteratively from the input of $\Phi_{[2][n-1]}(z)$ into the RHS of Eq. (6). The writing out of the summation of $\Phi_{[2][n]}(z)$ with respect to iteration index n is akin to the early methods of perturbative quantum electrodynamics done by Hans Bethe circa 1949[16], but only without the need to subtract infinities out. Renormalization methods are not required to solve Eqs. (6), analytically or computationally. The first iteration, $\Phi_{[2][1]}(z)$, contains logarithms of z, Ei functions of z, and terms of exp(-$\Sigma_t$ z) for several factors. The "2" refers to the 1-MeV neutrons and can be efficiently approximated by a much more tractable sum of a polynomials and log expressions. $\Phi_{[2][1]}(z)$ is the flux of fast neutrons which have been scattered only once. We allow batter-1 to refer to radiative particles which have undergone collisions only once. Following through with the iterations, $\Phi_{[2][2]}(z)$ is the flux of fast neutrons which have been battered twice. Likewise, $\Phi_{[2][3]}(z)$ is the flux of fast neutrons which have been battered three times. In general, $\Phi_{[m][n]}(z)$ is the flux of neutrons of group 'm' which have been battered n times without being either demoted from nor promoted from its energy-group "m". 'Fast' albedo is the ratio of the sum of fast scattering cross sections divided by the total fast cross section of neutrons. So long as the albedo is less than or equal to 1, the series of $\Phi_{[n][m]}(z)$ is guaranteed to converge. Five to eight iterations have proven to be required for sufficiently convergent solutions. We use the approximation that the current of escaping neutrons which have penetrated the wall is given by the sum of the 'unbattered' flux plus half (i.e., 0.5) of the sum of fluxes comprised of scattered neutrons, where 0.5 is the analytically and geometrically guaranteed minimum. Judiciously, we occasionally replace 0.5 with 0.52 or even 0.55, where this is the dimensionless factor of escape which relates surface $\Phi$ value to current density of scattered neutrons which depart from the surface.

For the sake of space and the context of this paper, we will focus on predictions for scattering deterministically of the transmission of neutrons through sample slabs of iron at various thicknesses and on the prediction for the portion of neutrons which are returned backwards (due to back-scatter) from the slab. Our deterministic method distinguishes between the population of fast scattered neutrons and of slow neutrons (where the choice of '1' for 'm' designates the down-scattered, or slow, neutrons). We also conducted Monte Carlo simulations of neutrons from a beam which approach the same wall of iron material. The two Monte Carlo (MC) codes used are: MCNP (which was developed and updated by LANL) and a custom Monte Carlo code which was developed early in 2014 [17,18] and had proven to be valid for the modeling of isotropic scattering. In Table (1) the name assigned to our custom MC code is SMUSKE, where SMUSKE is designed to simulate isotropic scattering or radiative particles in arbitrarily chosen rectangular geometries, as done in the past.

Three examples are given below for a broad rectangular slab of homogeneous cast-iron where a beam of neutrons approaches the barrier at normal incidence. Our deterministic predictions, the predictions of MCNP, and the predictions of the MC code, SMUSKE, are included in Table (1). In the common style, MFP shall denote Mean-Free-Path of neutrons in response to the total cross section for the fast (0.9 MeV to 1.0 MeV) neutrons. Our deterministic algorithm is named "IntegIterator", which stands for Iterative Integral Equation Solver. "Deterministic Iterator" is a slightly lengthy alias for our abbreviation of IntegIterator. As a reminder, Eqs. (6a) and (6b) are Fredholm integral equations in terms of mathematical structure. "Mfpm" in column one means the value of MFP/(Wall Thickness). "Grp#' in the 2$^{nd}$ column of the table refers to the energy group number of the neutrons. Grp# 2 includes all neutrons which are in the energy range of 0.9 MeV through 1.0 MeV. These are the designated fast neutrons. Grp# 1 denotes the slow neutrons, which include all the free neutrons which have energy in the range of 0.0 MeV through 0.160MeV. There were less than one in 5000 neutrons found in the energy range which lies in between the 'slow' Group and the 'fast' Group of neutrons, therefore, we do not include such an intermediate group of negligible population.

Table (1) **Summary of Forward Scattering and Back-scattering Predictions based on the Monte Carlo Codes of SMUSKE and MCNP and on the IntegIterator Deterministic Algorithm**.

| Mfpm | grp# | IntegItera backward | Smuske backward | Mcnp backward | IntegItera forward | Smuske forward | Mcnp forward |
|---|---|---|---|---|---|---|---|
| 1.0 | 1-slow | 0.0137306 | 0.0086246 | 0.10248209 | 0.012238 | 0.0083994 | 0.0859201 |
| 1.0 | 2-fast | 0.3356 | 0.31328 | 0.1379007 | 0.6374654 | 0.669173 | 0.6721823 |
| *Go to* | | *Mfpm* | *value of* | *one half* | | | |
| Mfpm | grp# | IntIter back | Smuske back | Mcnp back | IntIter fwd | Smuske fwd | Mcnp fwd |
| 0.5 | 1-slow | 0.0045931 | 0.0068838 | 0.0533270 | 0.0043261 | 0.0058638 | 0.0500006 |
| 0.5 | 2-fast | 0.202463 | 0.19875 | 0.0799309 | 0.788618 | 0.78625 | 0.816202 |
| *Go to* | | *Mfpm* | *value of* | *3/10* | | | |
| Mfpm | grp# | IntIter back | Smuske back | Mcnp back | IntIter fwd | Smuske fwd | Mcnp fwd |
| 0.3 | 1-slow | 0.008677 | 0.002346 | 0.03172595 | 0.007289 | 0.0023406 | 0.031896 |
| 0.3 | 2-fast | 0.130606 | 0.11286 | 0.0512400 | .863998 | 0.88179 | 0.884878 |

In the first third of Table (1) the thickness of the rectangular slab of iron is 1 MFP long, which is 2.884 cm. According to SMUSKE, out of 10,000 incoming fast neutrons, 3132.8 fast n's travel out or backward from the wall, and slow n's escape back toward the source of the beam. Accordingly,

SMUSKE predicts that 6,691.7 fast n's escape forwards out through the iron wall, and 84 slow n's escape forward. Comparing the Deterministic Iterator and SMUSKE's predictions starting with the thick wall at 2.884 cm first, we find that for the collective back-scattered n's, which escape out of wall backward, the percent differences between the Deterministic Iterator and SMUSKE are: 45.6% for slow n's (i.e., of grp.1) and 14.6% for fast n's (i.e., of grp.2).  For the rate of forward escape, or transmission, the percent differences between IntegIterator and SMUSKE are: 37% for slow n's (i.e., of grp.1) and -4.85% for fast n's (i.e., of grp.2).  We briefly consider the wall of 1.442 cm, for which Mfpm = ½, giving a rate of forward escape, or transmission, the percentage of disagreement between our IntegIterator and SMUSKE as: -30% for slow n's and 0.30% for fast n's (i.e., of grp.2).  We briefly consider the wall which is at 3 tenths of an Mfpm at 0.8652 cm, giving a rate of forward escape, or transmission, the percentage of disagreement between our IntegIterator and SMUSKE as: 103% for slow n's and -2.03% for fast n's (i.e., of grp.2).

Next, we compare the escapee numbers of SMUSKE to those of MCNP for n's: the rate of backward escape, returning to the beam source, when the wall is one MFP thick, the percentage of disagreement between the predictions of MCNP and SMUSKE are:  -168% for slow n's and 77% for fast n's.  For the rate of forward escape, when the wall is one MFP thick, the percentage of disagreement between the predictions of MCNP and SMUSKE are: -164% for slow n's and 0.448% for fast n's. For the rate of backward escape, returning to beam source, when the wall is ½ MFP thick, the percentage of disagreement between the predictions of MCNP and SMUSKE are:  -154.2% for slow n's and 85.2% for fast n's. For the rate of forward escape, when the wall is ½ MFP thick, the percentage of disagreement between the predictions of MCNP and SMUSKE are: -158% for slow n's and 0.3738% for fast n's. For the rate of backward escape, returning to the external beam source, when the wall is *3/10* of a MFP thick, the percentage of disagreement between the predictions of MCNP and SMUSKE are:  -172.4% for slow n's and 75.1% for fast n's. For the rate of forward escape, when the wall is *3/10* of a MFP thick, the percentage of disagreement between the predictions of MCNP and SMUSKE are: -172.6% for slow n's and 0.349% for fast n's.

On the other hand, on a more impressive note, there is only a -5.30% disagreement between the respective predictions of our Deterministic Iterator (i.e., IntegIterator) and those of MCNP for the number of forward transmitted fast neutrons through the wall which has 1 MFP (2.884 cm) of thickness.  Also, there is only a -3.437% disagreement between the respective predictions of IntegIterator and MCNP for the number of forward transmitted neutrons through the wall with thickness of ½ MFP (1.442 cm).   Likewise, there is a -3.738% disagreement between the predictions of SMUSKE and MCNP for forward escape when Mfpm= ½.  In addition, there is a -2.387% disagreement between the respective predictions of IntegIterator and MCNP for the number of forward transmitted neutrons through the wall with thickness of *3/10* of an MFP (0.8652 cm).

One can see that our deterministic IntegIterator makes predictions which are either as close as or almost as close in their respective predictions to those of SMUSKE to the predictions of forward escape and backward escape generated by using MCNP alone. IntegIterator and SMUSKE have comparable operation speeds if one is content with 1.5 percent statistical fluctuations of Monte Carlo from using SMUSKE. However, SMUSKE does not easily lend itself to decisive mapping of internal flux at high resolution. IntegIterator does, by default, offer the feature of internal mapping. We observe the effectiveness of SMUSKE and IntegIterator in spite of the disadvantage of using isotropic cross sections used in these codes. MCNP does consider angular probabilities in great detail for 'n'-scattering for all the well-known isotopes on the complete table of nuclides.

Indeed, in a nuclear historical context, the compilation of the 'scatter-kernel' data of MCNP was a project which spanned more than a decade. Thus, it would be very difficult to summarize such a large amount of angular data with approximations of the zeroth, first, and second order Legendre polynomials of the cosine of scatter-angle with any tractable and manageable database which could function without incurring "data strangulation" of an analytical iterator (e.g., our IntegIterator) written in Maplesoft, high-level Python, or similar language/package. Some ask, "why not just give the M.C. 'jobs' to GEANT4 to execute?". With all due acknowledgement of the formidable abilities of GEANT4, such as readily offering options of angular data streaming, the three smaller codes SMUSKE, MCNP, and IntegIterator are all faster and easier to work with than GEANT4 for designated rectangular walls of metal bombarded by n's.

**Conclusions**

Our deterministic iterative algorithm, IntegIterator, agrees reasonably well regarding the prediction of the energy distribution and the direction distribution to the corresponding distributions of energy and overall forward direction which are generated via our MCNP simulations, despite the deficiency of IntegIterator not being able to process anisotropic scattering in terms of the design of a 'scatter' kernel or helpful Green's Function.  This can be seen from the results posted in Table (1) and in the summary of Table (1) above. On the other hand, neither our deterministic code nor our isotropically designed SMUSKE agree extremely well with the predictions of back-scatter results from MCNP. All three methods of calculation agree well on the prediction of percent neutrons (sum of quasi-elastically scattered and down-scattered neutrons) which escape from the target slab of our ferrous metal informatively showing the trend of almost succeeding to sustain the population of free neutrons (albeit with records of past collisions and occasional energy reductions). In reference to "surviving" n's, 'free neutron' means neutron not captured by any atom. It is evident from the slowing down that every one of the iron slabs in Table (1) incidentally functions as a neutron moderator.  However, the directionality and ratio of down-scattering are subject to disagreement. As explained above, MCNP has extremely detailed cross section libraries.  Iron-56 turns out to be one of these extremely anisotropically scattering isotopes.  Apart from the work of Chandrasekhar [19] and his use of H-functions for tracing intensities of scattered photons, there is little analytical work recorded on predicting flux densities theoretically which are solutions to the MBTE in which the scattering cross sections are anisotropic. The discrete ordinate method used to solve the MBTE is impressive in its flexibility and is deterministic, but that method is not analytical - thus being completely numerical at every step, unlike our IntegIterator which offers some "term-by-term" perspective for the theorist.  It is 'somewhat' easy to do analysis of solutions of the MBTE when the neutron scatterers (i.e., the nuclei) are isotropic. Many experts of reactor physics and shielding analysis approximate the multi-group MBTE as a multi-group coupling of two, 3, or even 6 simultaneous diffusion equations of radiative particles. By its intrinsic nature, it is virtually impossible to do angle dependent "ray track tracing" of neutrons if one does modeling with a diffusion equation (or equivalently a pair of diffusion equations) rather than the MBTE.

It would be convenient from a clinical radiological treatment planners' point of view to carefully look up the data for B(E,z), where B(E,z) is the buildup factor included in Eq.(2). However, in regard to neutrons, buildup coefficient data either is not widely published or is not available to the broad national/international communities of health physicists or engineers.  Moreover, a significant benefit of our deterministic method (and algorithm) of IntegIterator is the superior speed which it offers by its retention of the definitions of chosen geometries per wall to be bombarded and in its output extraction compared to the time and duties required at the conclusion of a corresponding run of an MCNP simulation. Similarly using GEANT4 is often even more involved and tedious than MCNP.  Thus our 2-group IntegIterator algorithm and formulation is much faster than MCNP when predicting penetration ratios as well as the distribution of energy of penetrating neutrons.

The simulation of MCNP is sufficiently fast for modeling transmissions and down-scattering of neutrons through rectangular walls. However, the processing of the output data from the output files generated by the MCNP code require considerable data processing which is done best either in a UNIX console environment or a DOS console environment. Many of the younger physical engineers have little LINUX training and thus tend to rely on a Windows environment or a 'Mac-Windows' environment to process outputs of their chosen software modelers within Windows, 'Mac-Windows', or XWindows (if using a LINUX version of Maplesoft).  Maplesoft is the language which we have selected for IntegIterator in order to conduct our local Flux calculations. Our 'Maple' version of IntegIterator can operate within the environments of Windows, XWindows of Mac OS, and Linux - as valid versions of Maplesoft can be placed in these OS's.  Much of our Maplesoft code can be translated into Matlab code, for the accommodation of the preferred software environment of many electrical engineers for Matlab.  Another, great benefit of our deterministic code is that one can procure a polynomial approximation of the local dose of neutron flux at any depth within the metal. The structure of the spatial internal solution for flux is a combination of log and polynomial terms. With MCNP such a feat of mapping flux as a function of the depth in a wall would require updated writing of an Input file which is more than ten-fold more elaborate than the input file for the IntegIterator code for the same slab of metallic material. Internal depth profiling with SMUSKE also is a challenge, but less so than it is with standard MCNP input file declarations.

It is reasonable to anticipate a future effort of analysis of 3-group neutron flux distributions with respect to energy by radiation transport theorists. However, for now, we focus on constructing 2-group databases of neutron cross sections of various important materials besides just iron and boron and subsequently using the algorithm and code(s) of IntegIterator to predict collective forward escape and backward escape of neutrons which initially enter slabs of the respective materials of interest.